\documentclass[]{spie}

\usepackage{amsmath,amsfonts,amssymb}
\usepackage{graphicx}
\usepackage[colorlinks=true, allcolors=blue]{hyperref}

\usepackage{aas_macros}
\usepackage{xspace}
\usepackage{hyperref}

\newcommand{\simi}{\ensuremath{\sim}}

\newcommand{\degree}{\ensuremath{^{\circ}}\xspace}

\title{Six winters of photometry from Dome C, Antarctica: challenges, improvements, and results from the ASTEP experiment}

\author[a]{N.~Crouzet}
\author[b]{D.~M\'ekarnia}
\author[b]{T.~Guillot} 
\author[b]{L.~Abe} 
\author[b]{A.~Agabi} 
\author[b]{J.-P.~Rivet}
\author[b]{I.~Gon\c{c}alves}
\author[b]{F.-X.~Schmider}
\author[b]{J.-B.~Daban} 
\author[b]{Y.~Fante\"i-Caujolle}
\author[b]{C.~Gouvret}
\author[c]{D.D.R.~Bayliss} 
\author[d]{G.~Zhou}
\author[b]{E.~Aristidi} 
\author[e,f]{T.~Fruth} 
\author[e]{A.~Erikson} 
\author[e,g]{H.~Rauer} 
\author[h]{J.~Szul\'agyi} 
\author[b]{E.~Bondoux} 
\author[i]{Z.~Challita} 
\author[j]{C.~Pouzenc} 
\author[d]{F.~Fressin} 
\author[k]{F.~Valbousquet} 
\author[l]{M.~Barbieri}
\author[b]{A.~Blazit} 
\author[b]{S.~Bonhomme}
\author[i]{F.~Bouchy} 
\author[b]{J.~Gerakis} 
\author[m]{G.~Bouchez}

\affil[a]{Dunlap Institute for Astronomy \& Astrophysics, University of Toronto, 50 St. George Street, Toronto, Ontario, Canada M5S 3H4}
\affil[b]{Laboratoire Lagrange, Universit\'{e} C\^ote d'Azur, Observatoire de la C\^ote d'Azur, CNRS, Boulevard de l'Observatoire, CS 34229, F-06304 Nice Cedex 4, France}
\affil[c]{Observatoire de l'Universit\'e de Gen\`eve, 51 chemin des Maillettes, 1290 Sauverny, Switzerland}
\affil[d]{Harvard-Smithsonian Center for Astrophysics, 60 Garden St., Cambridge, MA 02138, USA}
\affil[e]{Institut f\"ur Planetenforschung, Deutsches Zentrum f\"ur Luftund Raumfahrt, Rutherfordstra{\ss}e 2, D-12489 Berlin, Germany}
\affil[f]{German Space Operations Center, German Aerospace Center, M\"unchener Strasse 20, D-82234 We{\ss}ling, Germany}
\affil[g]{Technische Universit\"at Berlin, Zentrum f\"ur Astronomie und Astrophysik, Hardenbergstra{\ss}e 36, D-10623 Berlin, Germany}
\affil[h]{ETH Zurich, Institut f\"ur Astronomie, Wolfgang-Pauli-Str. 27, 8093 Z\"urich, Switzerland}
\affil[i]{Aix Marseille Universit\'e, CNRS, LAM (Laboratoire d'Astrophysique de Marseille) UMR 7326, 13388 Marseille, France}
\affil[j]{Concordia Station, Dome C, Antarctica}
\affil[k]{Optique et Vision, 6 bis avenue de l'Est\'erel, BP 69, 06162 Juan-Les-Pins, France}
\affil[l]{Universidad de Atacama, Departamento de F\'isica, Copayapu 485, Copiapo, Chile}
\affil[m]{GEMAC, University of Versailles/CNRS, 45 av. des Etats-Unis, F-78035, Versailles Cedex, France}

\authorinfo{Further author information:\\N.C.: E-mail: crouzet@dunlap.utoronto.ca}

\begin{document} 
\maketitle

\begin{abstract} 
ASTEP (Antarctica Search for Transiting ExoPlanets) is a pilot project that aims at searching and characterizing transiting exoplanets from Dome C in Antarctica and to qualify this site for photometry in the visible. Two instruments were installed at Dome C and ran for six winters in total. The analysis of the collected data is nearly complete. We present the operation of the instruments, and the technical challenges, limitations, and possible solutions in light of the data quality. The instruments performed continuous observations during the winters. Human interventions are required mainly for regular inspection and ice dust removal. A defrosting system is efficient at preventing and removing ice on the mirrors. The PSF FWHM is 4.5 arcsec on average which is 2.5 times larger than the specification, and is highly variable; the causes are the poor ground-level seeing, the turbulent plumes generated by the heating system, and to a lower extent the imperfect optical alignment and focusing, and some astigmatism. We propose solutions for each of these aspects that would largely increase the PSF stability. The astrometric and guiding precisions are satisfactory and would deserve only minor improvements. Major issues are encountered with the camera shutter which did not close properly after two winters; we minimized this issue by heating the shutter and by developing specific image calibration algorithms. Finally, we summarize the site testing and science results obtained with ASTEP. Overall, the ASTEP experiment will serve as a basis to design and operate future optical and near-infrared telescopes in Antarctica.  
\end{abstract}

\keywords{Photometry, Exoplanets, Dome C, Antarctica}

\section{INTRODUCTION}
\label{sec: intro}

Exoplanets can be discovered and characterized through their transits, which occur periodically when the orbit of the planet crosses the stellar disk along the line of sight. This requires high precision photometry or spectroscopy. Dome C is located on the summit of a high plateau in Antarctica at an altitude of 3200 m, 1100 km inland, and offers exceptional conditions for such observations: the continuous night during the Antarctic winter allows extended time-series to be taken with almost no interruptions, and the atmospheric conditions are excellent with low winds, a dry atmosphere with low precipitations, a low sky background, and a high clear sky fraction. A French-Italian station, Concordia, has been built at Dome C and is fully operational since 2005. The station is inhabited thorough the year including by two astronomers during the winter. ASTEP (Antarctica Search for Transiting ExoPlanets\cite{Fressin2005}, PI: T. Guillot) is a pilot project that aims at detecting and characterizing transiting exoplanets from Dome C and to test this site for photometry in the visible. The project consists of two instruments: ASTEP South, a simple instrument consisting of a 10 cm lens that was sent to collect the first photometric data and to test the technical feasibility of such observations, and ASTEP 400, a 40 cm telescope designed and built by our team to perform high precision photometry under the extreme conditions of the Antarctic winter. In this paper, we focus on the operation of the ASTEP instruments, on the main technical challenges, and on the solutions that we implemented or could be adopted for future projects. We also summarize the main scientific results.

\section{LOGISTICS AT THE CONCORDIA STATION}

Concordia is a year round facility station which access is limited to the austral summer season 
due to the extreme weather conditions, as for all other Antarctic stations. For more than nine months 
per year, Concordia interacts with the outside world only through communication systems. There are two ways to access the station: 
\begin{itemize}
\item[-] by ship: from Hobart (Australia) to the coastal stations 
Dumont d'Urville (France) or Casey (Australia); or from Lyttelton (Christchurch's port, New-Zealand) to 
Mario-Zucchelli (Italy). 
\item[-] by intercontinental planes: from Hobart (Australia) to Casey station; or from Christchurch (New-Zealand) 
to Mario-Zucchelli (Italy) or Mc Murdo (USA).
\end{itemize}  

For transportation of people and selected light cargo, Twin Otter or Basler aircrafts are used and leave from the
coastal stations, while heavy equipment is brought to Concordia 
using ground traverses. Cap Prud'homme station (France, Italy), a small annex station on the continent 5\,km away 
from Dumont d'Urville, serves as the continental launching point for these inland traverses. A typical traverse convoy 
consists of two or three snow grading machines leading eight to ten tractors towing the cargo sledges, and 
covering 1\,200 km in 8-10 days depending on weather conditions. Approximately 150\,tons of cargo is brought
on each traverse, two thirds of which is fuel. 

Concordia station has, since 2013, a permanent Internet connection using a $\sim$4\,m antenna. The satellite 
connection allows us to take control of the ASTEP\,400 telescope at any time, from any location, either to change 
the observing programme or to update the control software. However, the bandwidth is not sufficient to enable huge 
data transfer. All around the station up to about 1\,km, scientific instruments are connected by fiber optics 
to a switch located in the main buildings, where the network runs at 1\,Gbps.

\section{OVERVIEW OF THE ASTEP INSTRUMENTS}

\subsection{ASTEP South}

ASTEP South is a 10 cm refractor equipped with a 4k $\times$ 4k CCD camera in a thermalized enclosure, pointing continuously towards the celestial South pole \cite{Crouzet2010} (Figure \ref{fig: instruments}). The box is closed by a double glass window filled with nitrogen to ensure thermal insolation and prevent frost between the windows. This simple design with no moving parts enabled a quick installation at Dome C, allowed us to acquire the first photometric data from 2008, and served as a pathfinder for the larger 40 cm telescope. The field of view is $4^{\circ} \times 4^{\circ}$ with a pixel size of 3.4 arcsec/pixel and contains 6000 stars up to magnitude V = 16. The exposure time is 30 seconds and the PSF (Point Spread Function) FWHM (Full Width Half Maximum) is nominally 2 pixels. We use a R-band filter to minimize the sky background. With this design, the stars rotate around the CCD every sidereal day: they move by 4.5 pixels per exposure at the edges of the field and their PSFs are elongated. ASTEP South observed continuously for 4 winters from 2008 to 2011 and collected 870 000 science images. Lightcurves were built for 6 000 stars from magnitude 5 to 16 in the R band.

\subsection{ASTEP 400}

\begin{figure}[htbp]
   \centering
   \includegraphics[width=6cm]{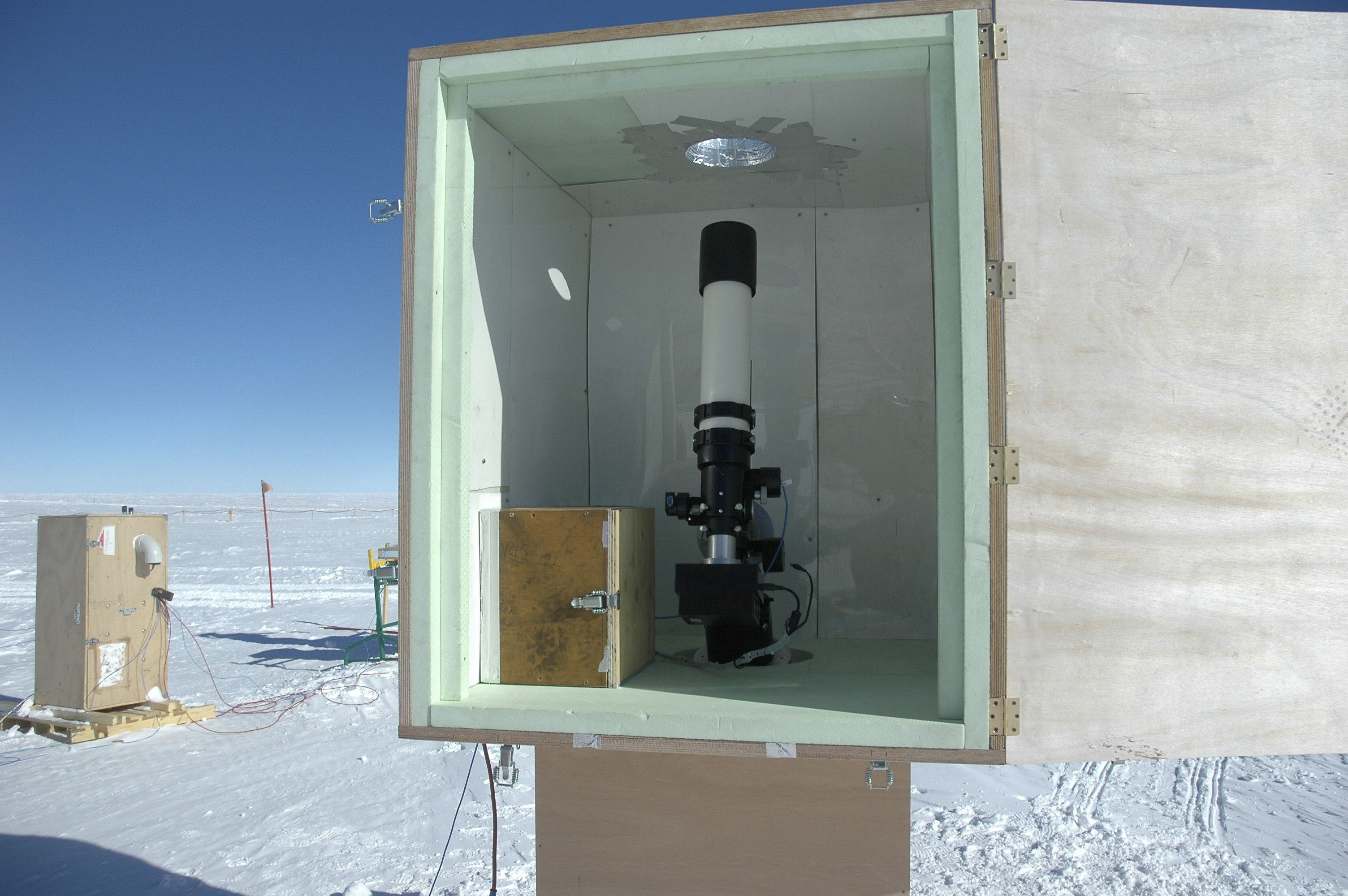}
   \hspace{2cm}
   \includegraphics[width=6cm]{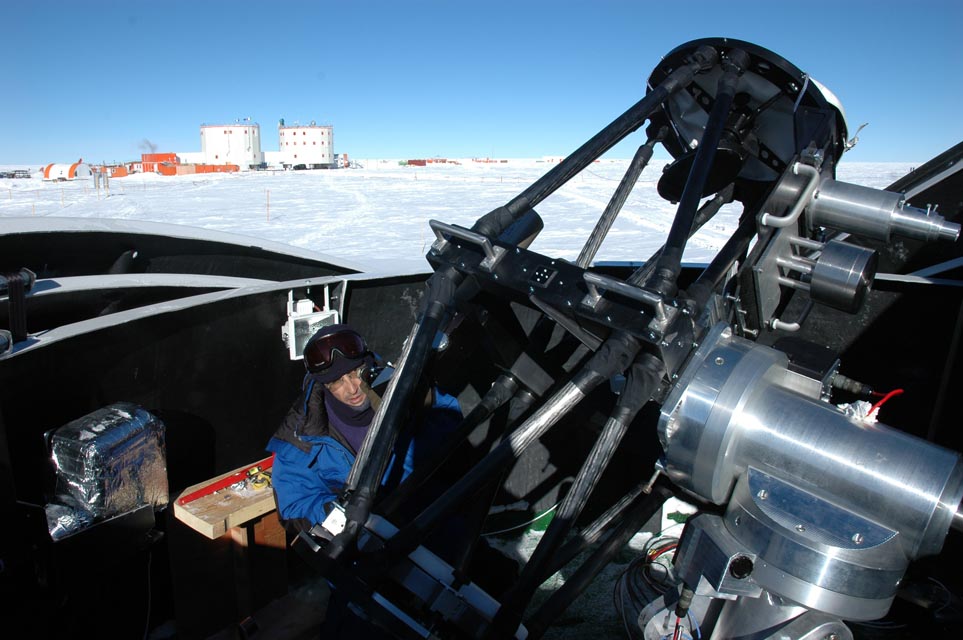}
   \caption{ASTEP South (left) and ASTEP 400 (right) installed at the Concordia station, Dome C, Antarctica.}
   \label{fig: instruments}
\end{figure}

ASTEP 400 is a 40 cm Newton telescope designed and built at the Observatoire de la C\^{o}te d'Azur to perform high precision photometry under the extreme conditions of the Antarctic winter \cite{Daban2010} (Figure \ref{fig: instruments}). Indeed, during the winter at Dome C, the ambient temperature varies between -30 and -80{\degree}C with variations up to 20{\degree}C in a few hours, which require specific materials and design. The telescope has a f/4.6 focal length with a 40 cm primary mirror made in Zerodur$\rm^{TM}$, and a field of view of $1^{\circ} \times 1^{\circ}$. The structure is composed of aluminium alloy parts joined by carbon fiber bars and is designed to minimize mechanical deformations due to temperature fluctuations. A focal box is attached to the structure and has two thermalized compartments. The upper one only includes optical elements and is stabilized to -40$^{\circ}$C. The lower one contains the cameras, their electronics and a MICOS piezoelectric actuator. Because of material requirements, this compartment is heated to -10$^{\circ}$C. Furthermore, because of problems encountered at low temperatures, the shutter of the science camera was heated to +5{\degree}C (see Section \ref{sec: shutter issues}). The science camera is a FLI Proline 4k $\times$ 4k pixels with 0.9 arcsec/pixel, and the guiding camera is a SBIG. A dichroic beam-splitter sends the red and blue wavelengths to the science and guiding camera respectively, resulting in a science bandpass from 600 to 800 nm. The total weight of the structure and camera box is 100 kg. The mount is equatorial (model AstroPhysics AP-3600), and the ensemble is placed inside a dome. ASTEP 400 observed for 4 winters from 2010 to 2013 and observed 22 different fields of view. More than 200 000 frames were recorded and 310 000 stars were analyzed.

\section{OPERATION OF ASTEP}

\subsection{Control and Acquisition Software}

Both ASTEP South and ASTEP 400 were driven by custom-made software programs designed to operate both in a manual or automatic mode. The ASTEP South software controlled the acquisitions, ran a predefined observing sequence taking into account the Sun's elevation, transferred the data from the shelter next to the instruments to the station, and saved them onto hard drives.
The ASTEP 400 software is fed by a custom script describing a set of independent observing fields. Each field has its own observational constraints (coordinates, priority, temporal boundaries, sun elevation, periodicity...) which are processed by the software in order to schedule the observations. To ensure proper pointing of the telescope in automatic mode, the software uses a built-in field recognition algorithm capable of repointing the telescope even in the case of a pointing error that is larger than~1$^{\circ}$.
Both programs included an automated daily data quality check based on an analysis of the raw images of the previous day: we compute the number of stars present in the field of view as a function of time, the mean intensity of the images, the PSF variations, the ambient and various system temperatures, and the pointing variations. A summary file is generated and sent by email to the team in Nice. This daily feedback is used to monitor the behaviour of the instruments and to prepare human interventions when necessary.

\subsection{Data Storage} 

ASTEP 400 is connected to a dedicated computing and mass storage server (48 CPUs, \simi20 TiB). Data are locally stored onto a data acquisition PC (2 TiB disk capacity) and backed up automatically, every day, on the mass storage. A second backup is made on TiB-capacity hard drives and sent back to our laboratory in Nice during the summer campaign. Another backup on hard drives is kept on-site for security in the (rare) cases of disk failures or shipping problems. In 2013, a permanent Internet link allowed us to take control of the telescope at any time and from any location, either to change the observing programme or to update the control software. However, the bandwidth is not sufficient to enable the transfer of the science images. The ASTEP South data are also saved using a dedicated PC and stored on hard drives in Concordia, and sent to Nice using hard drives.

\subsection{Human Interventions}

For ASTEP South, human interventions occurred at the beginning of each winter. The instrument's enclosure was kept thermalized during the summer months but was not monitored continuously as during the winter. As a result, the pointing of the South pole and the focusing had to be adjusted at the beginning of each campaign. After this initial setup, human interventions were rare and risky during the winter. In 2008, several interventions occurred during the first two months. In 2009, one major intervention occurred around mid-winter to fix the thermalization and install a baffle around the entrance window. In 2010 and 2011, no major intervention was necessary. Minor interventions occurred after winter storms to clean the baffle and the entrance window from the deposited ice dust.

ASTEP 400 is a semi-robotic and fully computer-controlled instrument, so that very few human interventions are required. Only regular inspections of the primary and secondary mirrors are needed. Through a single graphical user interface, the winter-over crew member in charge of ASTEP 400 can monitor the relevant parameters of the telescope, perform any modification, send orders to the telescope motor drives, and run pre-determined observation scripts from the main Concordia building. Minor cleaning of the telescope was necessary after winter storms.

\section{MIRROR DEFROSTING}

During the Antarctic winter, the ambient temperature can fall down to -80{\degree}C and is characterized by rapid variations of several degrees per hour. In such conditions, the relative humidity leads to ice nucleation and frost formation on the primary mirror of ASTEP 400 and, to a lesser extent, on its secondary mirror. Ice dust can also fall on the primary mirror. Blowing dry air (continuously or from time to time) on optical surfaces or warming them up prevents the frost formation and/or snow deposit.
We have installed, during the 2010-2011 summer campaign (the second year of operation of ASTEP 400), a defrosting setup for both the primary and secondary mirrors. This system consists in a custom designed film resistor with electrical properties summarized in Table \ref{tab: defrosting resistors}, attached to the rear surface of each mirror. Temperature probes were glued on the side of the mirrors and a pair of additional temperature controllers was added to drive the heaters. This way, the temperatures of the mirrors could be monitored easily, and the power supplied to the heaters was adjusted with the aim of minimizing turbulent plumes inside the telescope.
This defrosting setup can operate both in ``preventive'' or ``curative'' mode. In the preventive mode, a small fraction of the maximum power is supplied to the heaters, so as to maintain the mirrors' surface temperature above the ambient temperature (an example is given in Table~\ref{tab: defrosting example}). In the curative mode, 100\% of the nominal power of the heaters is supplied for a short period to remove the frost and ice dust deposit. This mode is activated under special weather conditions and no science images are recorded during this phase.

\begin{table}
\begin{center}
\caption{Electrical properties of the defrosting resistors. The primary and secondary mirrors are noted M1 and M2.}
\label{tab: defrosting resistors}
\vspace{3mm}
\begin{tabular}{lll}
\hline
\hline
 & M1 & M2 \\
\hline 

Resistance [$\Omega$] &	2.3	 &	5 \\
Intensity under 24 V [A] &	10.4 &	4.8 \\
Power under 24 V [W]  & 	250 &	115 \\
Surface power [W/cm$^2$] &	0.9 & 	0.41 \\

\hline
\hline
\end{tabular}
\end{center}
\end{table}

\begin{table}
\begin{center}
\caption{Example of power applied to the defrosting resistors and resulting mirror temperatures $T_M$ for an ambient temperature $T_{amb} = -23{\degree}$C in the preventive mode. Column~4 is the power relative to the maximum power and column~5 is the absolute power. The primary and secondary mirrors are noted M1 and M2.}
\label{tab: defrosting example}
\vspace{3mm}
\begin{tabular}{ccccc}
\hline
\hline
 &  $T_M$ [{\degree}C] & $T_M - T_{amb}$ [{\degree}C] & Power [\%] & Power [W] \\
\hline 

M1	&	$-14$    &	 $+9$	&	5	 &	12.5 \\
M2	&	$-10$	&	$+13$	&	5	 &	5.8 \\

\hline
\hline
\end{tabular}
\end{center}
\end{table}

\section{PSF STABILITY}
  
\begin{figure}[htbp]
   \centering
   \includegraphics[width=7cm]{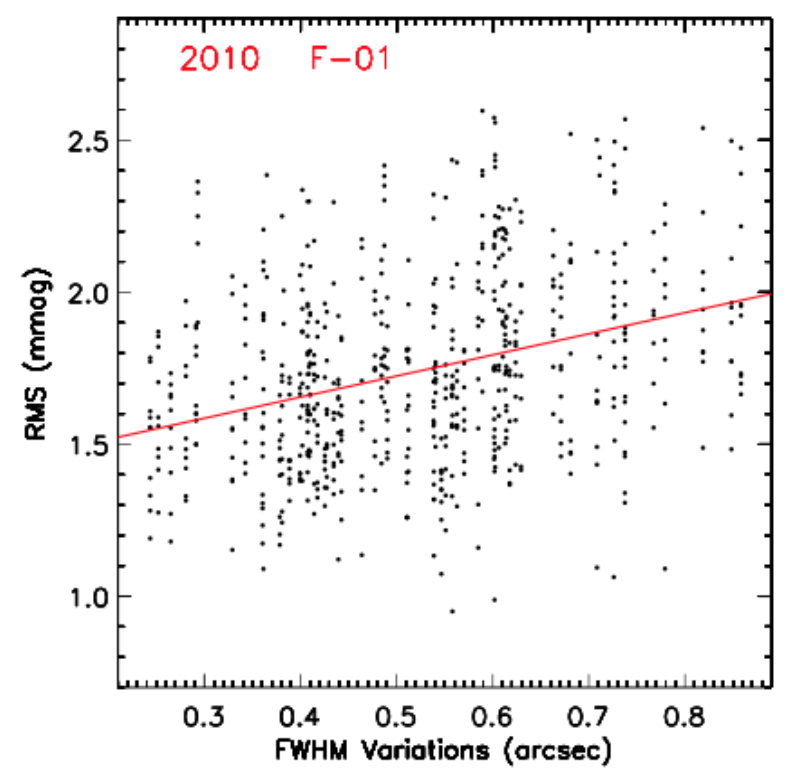}
   \caption{Photometric RMS of the lightcurves of selected bright stars observed by ASTEP 400 in 2010 as a function of the night-to-night variation of the PSF size, for the field of view denoted F-01. The positive correlation is indicated by a red line.}
   \label{fig: rms fwhm ASTEP400}
\end{figure}

ASTEP 400 is the first optical telescope to have functioned properly during four years in the extreme conditions of the Antarctic plateaus. In spite of very promising results, we were not able to reach the telescope's nominal photometric precision. The PSF size was on average 4.5 arcsec instead of the 2 arcsec nominally envisioned. This large and variable PSF size led to a lower photometric accuracy for faint stars because of crowding \cite{Fruth2014}, as well as for bright stars as illustrated in Figure \ref{fig: rms fwhm ASTEP400}. It also led to difficulties in processing the data for dense fields, in particular when following microlensing events towards the galactic center. This correlation between the PSF FWHM and the atmospheric seeing is also evident in the ASTEP South data \cite{Crouzet2010}.

To understand how rapid temperature variations could affect the photometry, we calculated thermal deformations of the ASTEP 400 telescope and analyzed their expected consequences on the turbulence along the optical path \cite{Guillot2015}. We also conducted a deep analysis of the data and performed specific tests during the summer campaign to identify the influence of the various subsystems on the image and photometric quality.
We identified the causes of these large PSFs. First, ASTEP 400 is installed at only 2 m above the ground and suffers from the large (\simi2 arcsec) and variable atmospheric seeing. Second, the primary mirror was heated to avoid frost and was generally 5 to 10{\degree}C above the ambient temperature. This was generating turbulence by free convection and was responsible for a seeing of the order of 1 to 2 arcsec. Third, the focus of the camera was tested infrequently, and is also partly responsible for the PSF broadening. Another source of broadening is some astigmatism that was due to an imperfect alignment of the optics.

In order to minimize the photometric noise and reach a higher spatial resolution, we suggest the following improvements:
\begin{itemize}
\item The mirrors should be heated only slightly above the ambient temperature. Rather than the 5-10 K temperature difference used essentially out of convenience by the winterover observers at Concordia, we estimate that ensuring a mirror temperature only 1-2 K above the ambient would yield a modest mirror seeing of \simi0.4 arcsec \cite{Guillot2015}. This can be best achieved with a new system to warm only the reflective part of the mirror.
\item The telescope should be kept in focus. This can be easily achieved by monitoring the dependence of the focal position with the temperature(s) and with regular tests, in order to build an instrument model that can be used within a feedback loop.  
\item A fine alignment of the optical elements is necessary. This implies redesigning the optical box so it includes a laser that could be used to perform the alignment without having to remove elements from the box.
\item Atmospheric seeing and its consequences should be minimized. This can be done by using a tip-tilt system and by putting the telescope 8 to 12 meters high to avoid most of the turbulent layer.
\end{itemize}

\begin{figure}[htbp]
   \centering
   \includegraphics[width=7cm]{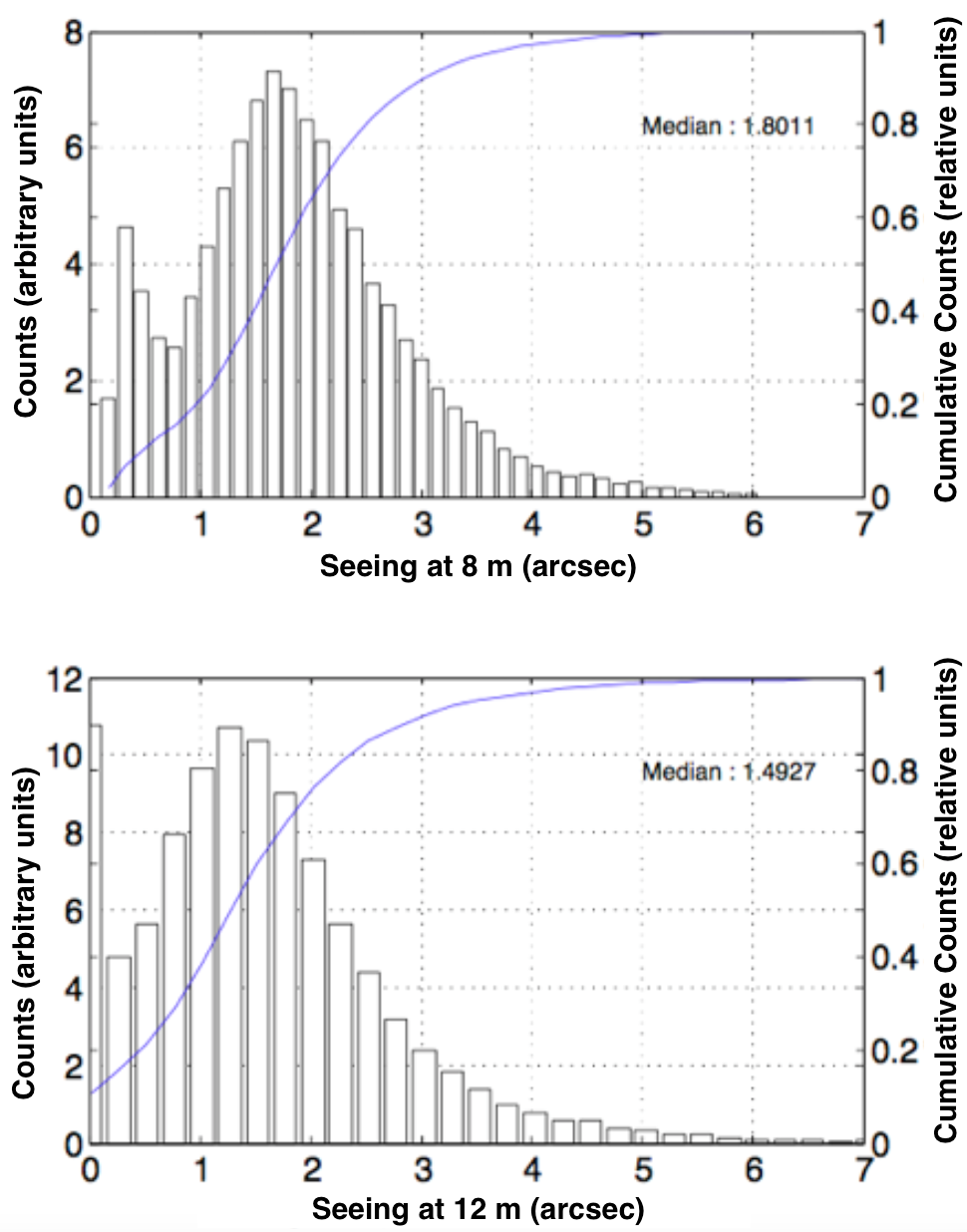}
   \caption{Histograms of the typical atmospheric seeing at Concordia, Dome C, at an elevation of 8 m (top) and 12 m (bottom). The cumulative count is indicated as a blue line. The median seeing at the elevation of ASTEP (2 m) was 2.3 arcsec. Poor seeing conditions ($>$ 2 arcsec) drop to 40\% of the time at 8 m and to 25\% of the time at 12 m elevations.}
   \label{fig: seeing histograms}
\end{figure}

Concerning the latter point, we estimate that the telescope should be installed at an elevation of 8 meters or more. This elevation corresponds to that of the current wooden platforms installed at Concordia which unfortunately cannot be used directly because of vibrations. ASTEP 400 was installed on a small concrete pillar only \simi2 m above ground, implying an estimated median seeing of 2.3 arcsec. Moving the telescope to 8 m would decrease this value to 1.8 arcsec. At a 12 m elevation, this value would decrease further to 1.5 arcsec, and more importantly, seeing conditions over 2 arcsec would occur only 25\% of the time \cite{Aristidi2009, Fossat2010, Aristidi2013, Petenko2014} (Figure \ref{fig: seeing histograms}).

\section{ASTROMETRIC PRECISION}

\subsection{Field Matching and Pointing Variations of ASTEP South}

The ASTEP South instrument is fixed, with no guiding or tracking system. The pointing must be adjusted manually using the mount and consists in placing the celestial South pole at the centre of the CCD, nominally at pixel 2048, 2048 $\pm$ 50, 50. This pointing is done at the beginning of each winter. First, an initial pointing within 2 degrees of the South pole is reached by taking exposures of a few minutes, analyzing visually the PSF elongation pattern, and adjusting the mount. Second, a fine pointing is required. To facilitate this, we developed a custom-made field matching algorithm that identifies the position of the South pole on the CCD. This program runs from the acquisition PC located in the shelter next to the instrument, and is launched on individual images immediately after they are taken. The position of the South pole is given in pixel units and is used by the winter over crew to adjust the mount. Using this procedure, a nominal pointing is reached after a few iterations.
The field matching algorithm runs as follows. The target star coordinates extracted from the TYCHO2 and UCAC3 catalogs are converted into pixel units. Point sources in the image are identified using the FIND procedure from the DAOPHOT package. We find the geometrical transformation consisting of a translation and a rotation that yields the best match between both sets of coordinates. Then, we use the inverse transformation to compute the location of the South pole on the image (we convert the coordinates 0\degree, -90\degree into a pixel location). The typical precision of this algorithm is 0.2 pixels. In the main data reduction pipeline used to extract the lightcurves, we also correct for distortions and we reach a precision of 0.1 pixel on the location of the target stars.
We investigate the mechanical stability of ASTEP South by measuring the motion of the celestial South pole on the CCD over the winter. Within one day, the South pole is stable within 1 pixel. Over 100 days, the South pole varies typically by 18 and 9 pixels in the $x$ and $y$ directions respectively. This long term motion is oriented along the North-South direction which is the vertical axis of the instrument, and may correspond to a bending of the instrument under its own weight or to a motion of the ice. Overall, ASTEP South is very stable during the winter.

\subsection{Guiding of ASTEP 400}

Tracking defects producing noticeable shifts of the stellar PSFs on the science CCD sensor are likely to yield photometric noise. Thus, special efforts have been made at the design level to make the tracking as accurate as possible. For this sake, a guiding camera has been inserted in the thermally controlled camera box, and receives the $400-550$~nm band transmitted by a dichroic beam-splitter (the $600-800$~nm band is reflected to the science camera). A software real-time regulation loop detects tracking defects and sends the appropriate corrections to the motors' drives.
To assess the efficiency of this system, we have selected a sample of $3885$ frames from the science camera taken in May 2011 over 6 nights on the same field (center of field\,: star UCAC4\,172-184045). For each frame, an astrometric reduction was performed using a local implementation of the ``astrometry.net'' on-line astrometry program ({\tt http://astrometry.net/} \cite{Lang2010}). This yields the J2000 coordinates of the central pixel, the on-sky field orientation, and the plate scale.
On the sample under study, the on-sky scale fluctuates between $0.9295$ and $0.9297$~arcsec/pixel. This slight fluctuation yields only sub-pixel PSF motion, even in the corners of the image.
The rotation angle has a sinusoidal time-variation with an amplitude of $7$~arcmin. This is a consequence of a polar misalignment, due to both slight initial alignment defects and to long-term ice creeping. For pixels close to the corners of the image, this leads to PSF motions of $\pm6$~pixels.
In addition, the central pixel has a residual on-sky motion but stays within an error box of $\pm2$~pixels in declination and $\pm7$~pixels in right ascension. This is due to internal flexures that shift the guiding camera with respect to the science camera.
Position/orientation fluctuations would deserve to be corrected for the next generation of instruments. This would require extra engineering efforts to lower internal flexures, to stabilize the polar alignment, and/or to introduce automated compensations of the on-sky image shifts and rotations.

\section{IMAGE CALIBRATION}

\subsection{Flat-Fields}

\begin{figure}[htbp]
   \centering
   \includegraphics[width=8cm]{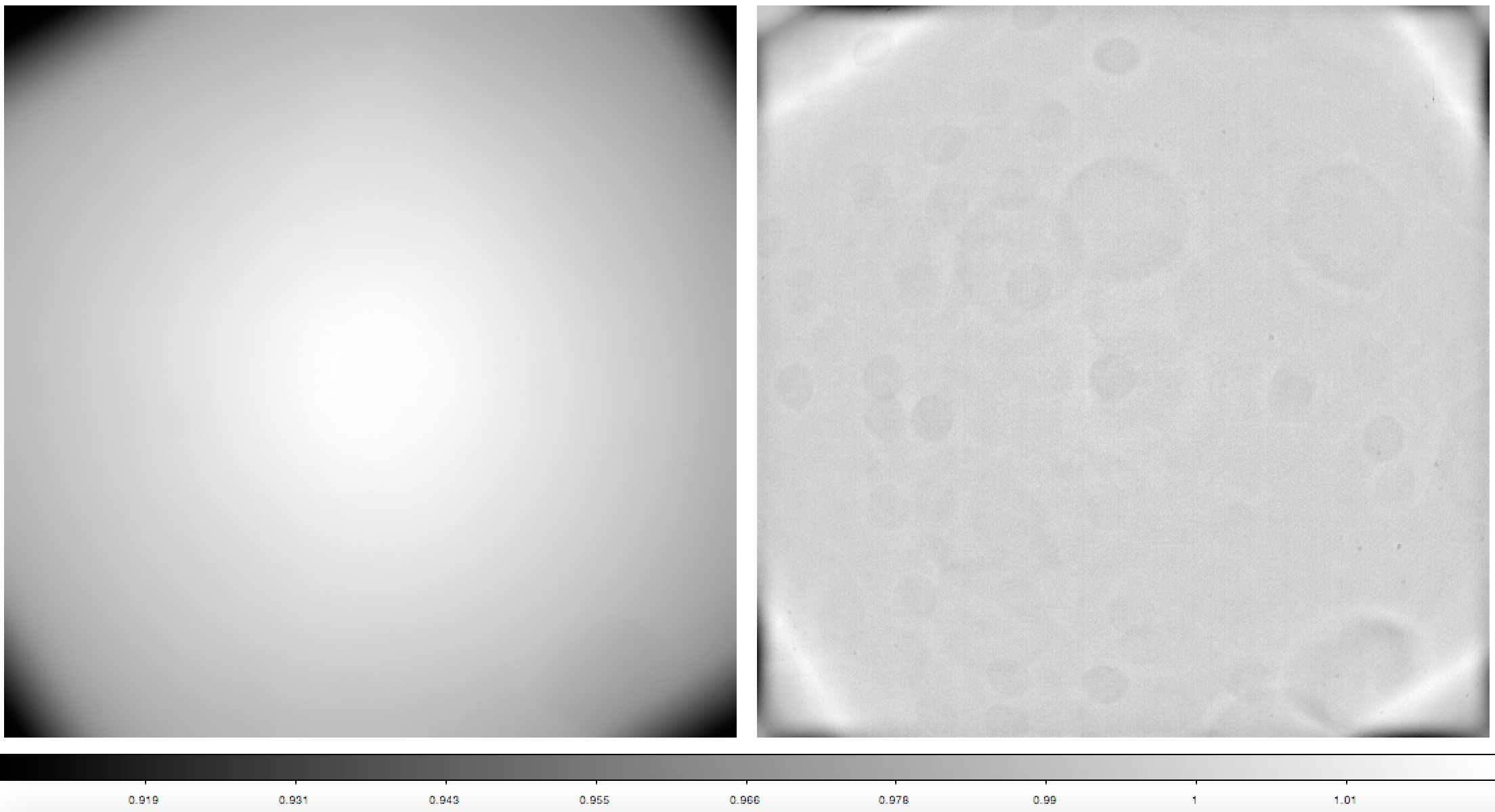}
   \caption{Left: Flat-field obtained for ASTEP South for the 2009 season. Right: Same flat-field after subtracting the low frequency component. The scale in ADU is displayed at the bottom.}
   \label{fig: flat field}
\end{figure}

In temperate sites, flats-fields are usually taken at twilight at the beginning or end of the night. At Dome C, the sky background increases everyday around noon. This is invisible to the naked eye but is evident in the science images. As the Sun elevation increases, this illumination quickly saturates the images (except in June). We use the images with a background that is below half of the CCD dynamic range to create a flat field. For ASTEP~South, the stars rotate around the CCD and we use interleaved medians to remove point sources. For ASTEP~400, we apply a small motion to the telescope to shift the point sources and also apply median filters. This technique yields high quality flat fields as shown in Figure~\ref{fig: flat field}.

\subsection{Shutter Issues}
\label{sec: shutter issues}

\begin{figure}[htbp]
   \centering
   \includegraphics[width=11cm]{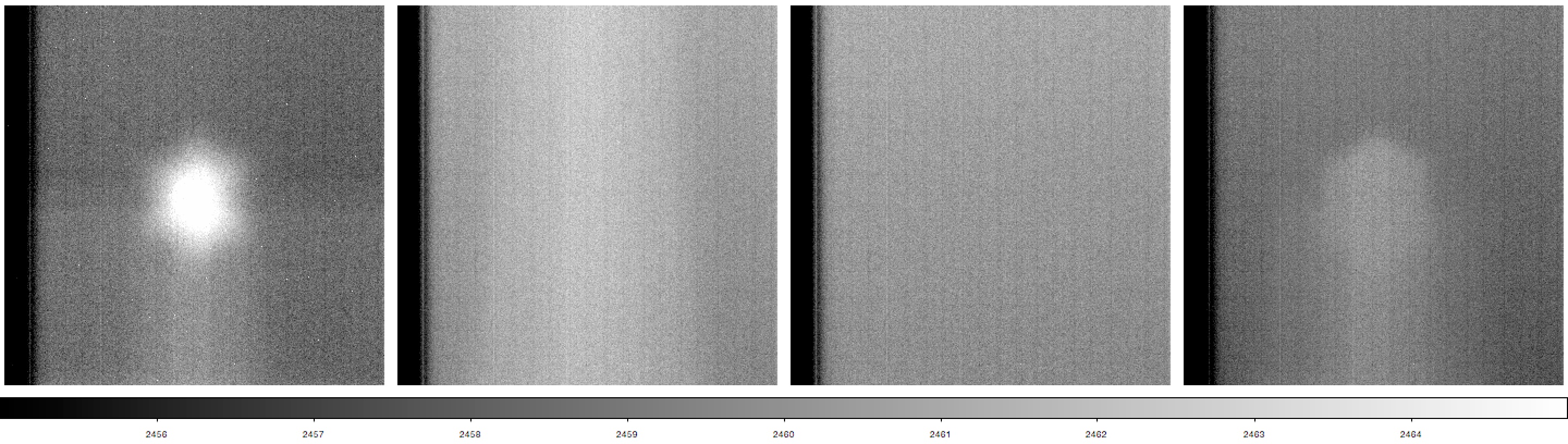}
   \hfill
   \includegraphics[width=5.6cm]{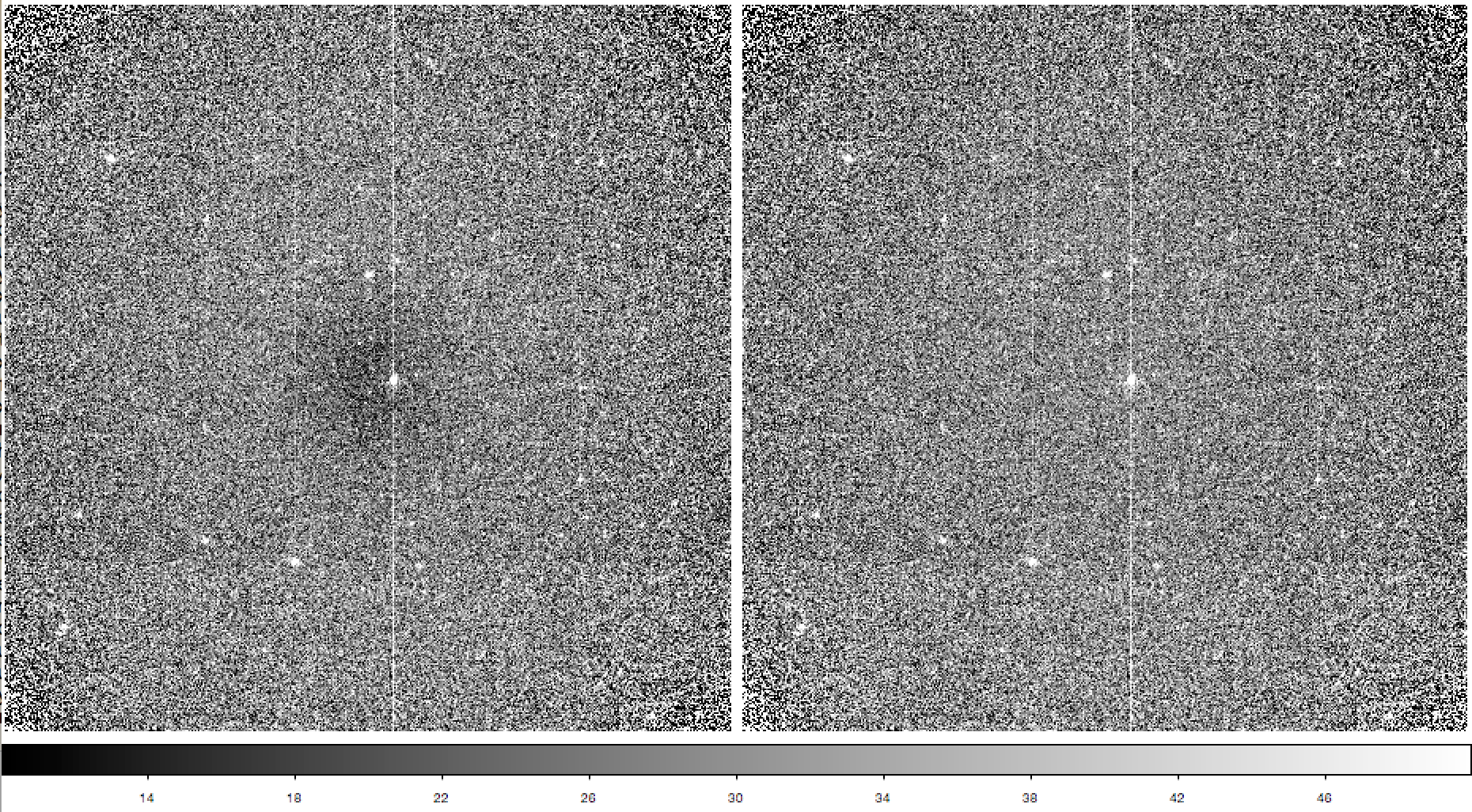}
   \caption{Examples of images from ASTEP South taken with a defective camera shutter. Left to right: Masterdark, masterbias, masterbias corrected from the shutter effect, final calibration image, and example of a science image calibrated with the masterdark and with the final calibration image. The scale in ADU is displayed at the bottom.}
   \label{fig: shutter issues}
\end{figure}

A major issue we encountered was related to the camera shutter (the camera is a ProLines series from Finger Lake Instrumentation). During the setup of ASTEP South at the beginning of the first winter, we found that the shutter did not close properly at negative temperatures, whereas the enclosure was thermalized to -20{\degree}C to minimize the difference with the ambient temperature and the temperature of the camera electronics. We maintained the shutter at a temperature of +5{\degree}C by taping resistor films around it. The shutter worked nominally for almost two winters. Towards the end of the second winter, after \simi~393 000 open-close cycles, it started malfunctioning again. The bias and dark images were strongly affected and could not be used directly to calibrate the science images (Figure \ref{fig: shutter issues}). We handled this issue in two ways. First, we heated the enclosure to a much higher temperature, between +25 and +30{\degree}C; the shutter still did not close completely. Second, we created a specific algorithm to build a usable calibration image from the affected bias and darks. The quality of the lightcurves obtained after these corrections is similar to that obtained when the shutter was working properly. The ASTEP~400 camera is the same model as that of ASTEP South and we encountered similar issues with the shutter.
~\\

\section{SUMMARY OF THE SITE TESTING AND SCIENCE RESULTS}

\begin{figure}[htbp]
   \centering
   \includegraphics[width=4.2cm]{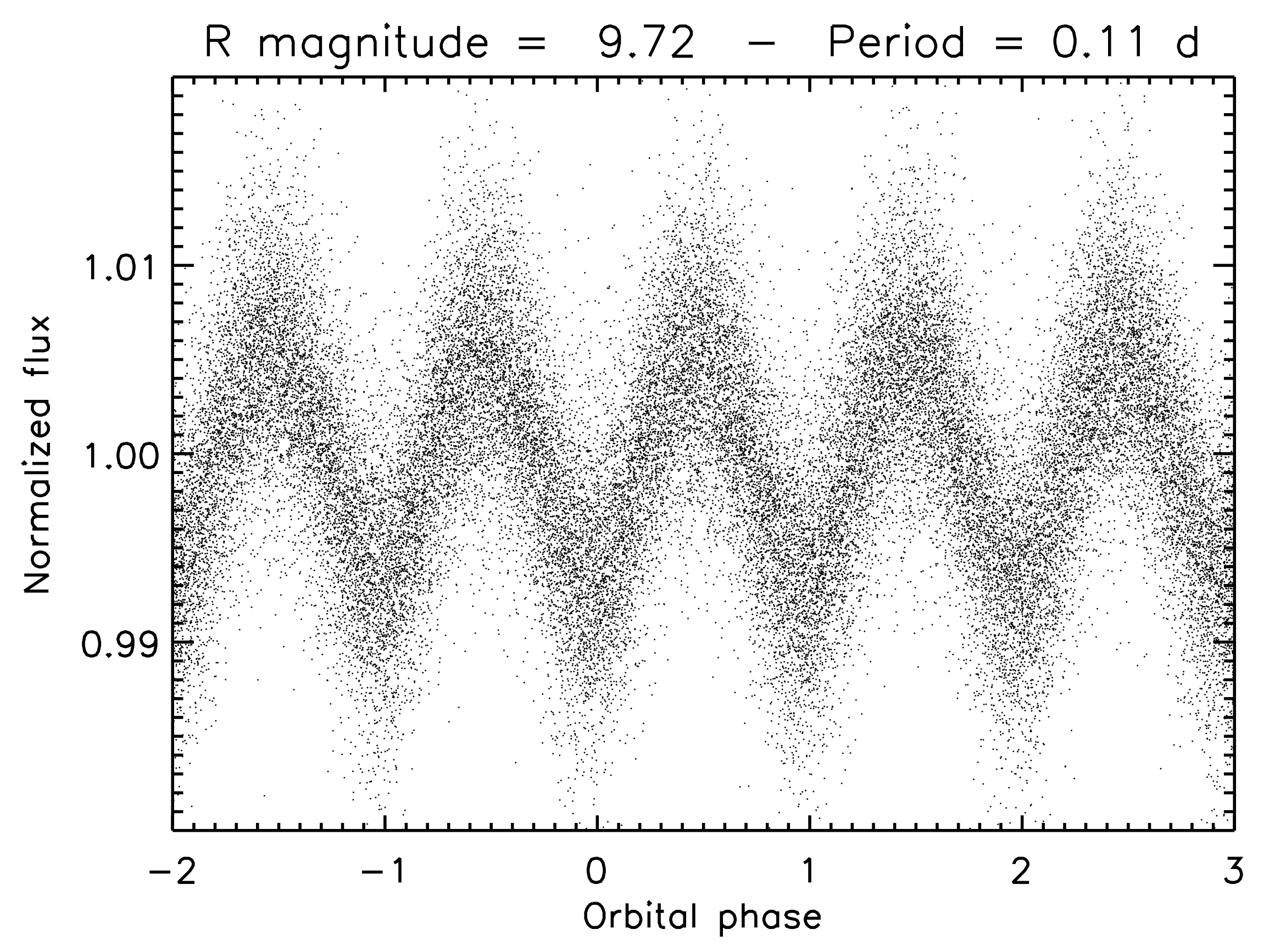}
   \includegraphics[width=4.2cm]{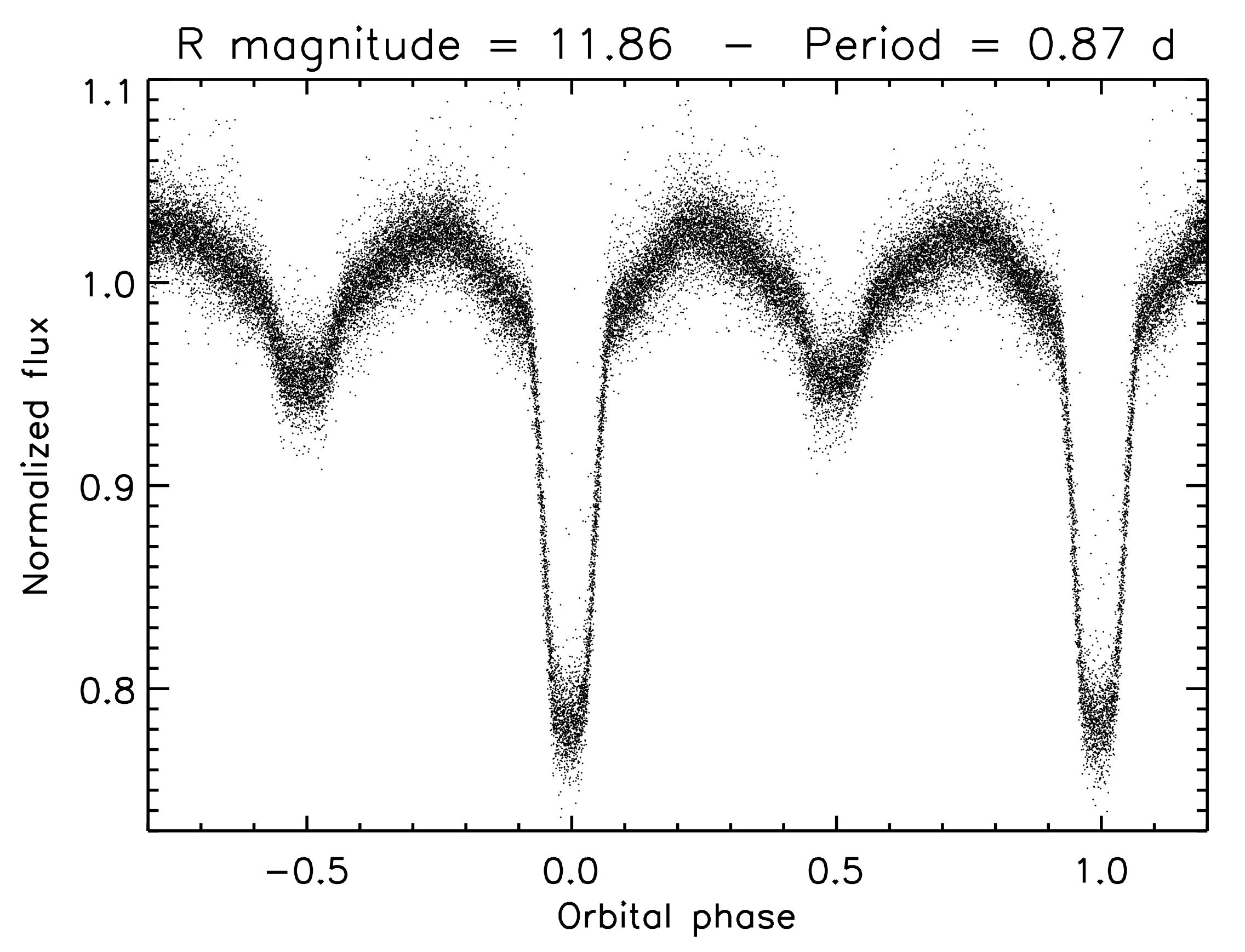}
   \includegraphics[width=4.2cm]{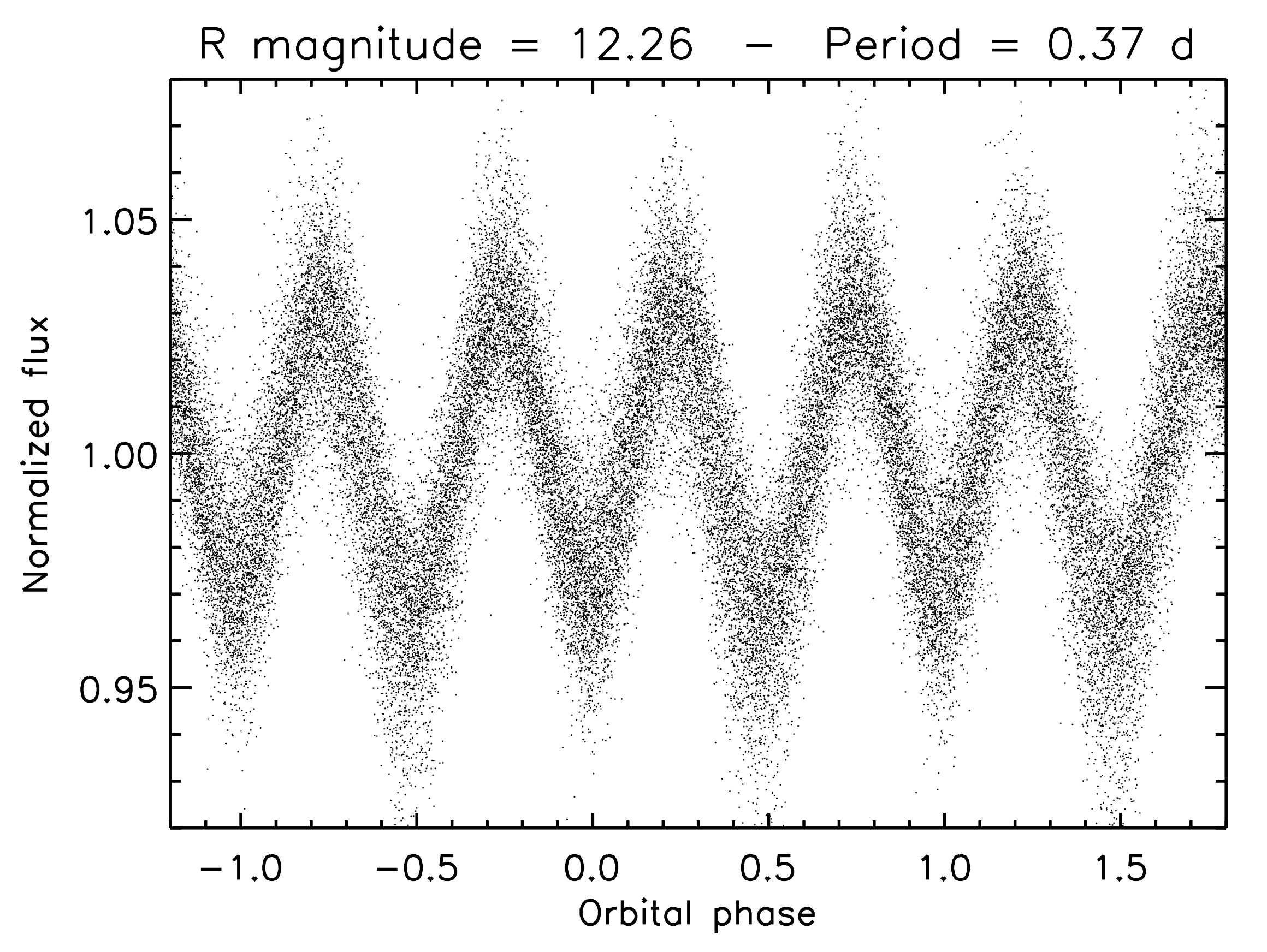}
   \includegraphics[width=4.2cm]{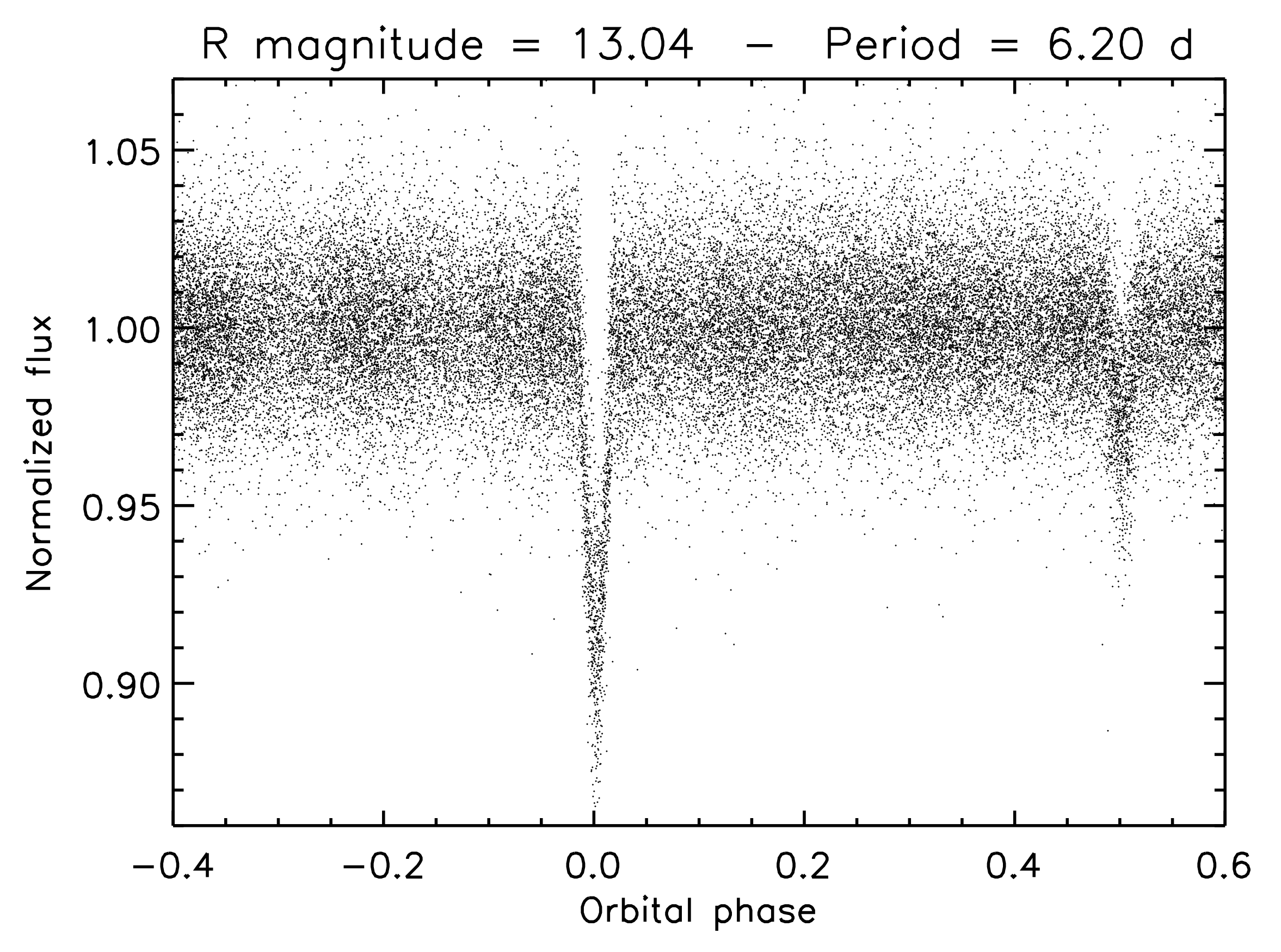}
   \caption{Examples of lightcurves of variable stars and eclipsing binaries obtained with the 10 cm instrument ASTEP South.}
   \label{fig: ASTEP South lightcurves}
\end{figure}

The scientific and site testing results obtained with ASTEP are detailed in other papers. Here, we present a brief summary of the main results. With ASTEP South, we assessed the feasibility of continuous photometric observations from Dome C, and quantified the effects of the Sun and the Moon. Although not detected with the naked eye, the Sun increases the sky background a few hours per day and affects the photometry when its elevation is higher than -13\degree. This effects is not significant in June, but 3 to 7 hours per day are lost in May and July, and 8 to 12 hours per day are lost in April and August. The full Moon affects the photometry as in any other sites, and this illumination increases under veiled or cloudy conditions due to reflection by clouds. We also quantified the clear sky time fraction during first winter, and found between 56.3\% and 68.4\% of excellent weather, 17.9 to 30\% of veiled weather during which stars are still visible, and 13.7\% of bad weather\cite{Crouzet2010}. The next three seasons as observed with ASTEP South yield similar or better results. For ASTEP 400, the effective duty cycle over three winters including bad weather and technical failures is 65.14\%\cite{Mekarnia2016}, in line with ASTEP South. 
With ASTEP 400, we detected the secondary eclipse of an exoplanet in the visible from the ground for the first time: we observed the hot Jupiter WASP-19b during one month and measured an eclipse depth of $390\pm190$ ppm, which is unprecedented for a 40 cm telescope on the ground. We derived a corresponding brightness temperature of $2690\pm200$~K in the 600-800 nm range, allowing us to constrain the temperature profile and circulation in the planet's atmosphere \cite{Abe2013}. In addition, we detected 43 transiting planet candidates\cite{Mekarnia2016}, 1156 variable stars, and 674 eclipsing binaries\cite{Chapellier2016}. With ASTEP South, we detected a few hundreds of variable stars (examples are shown in Figure \ref{fig: ASTEP South lightcurves}); we also detected a long-period stellar eclipsing binary in which the companion is a low-mass star: the orbital period is 75.58 days, the eccentricity is $0.260 \pm 0.007$, the primary star is a G9, and the companion has $M_{sec} = 0.229 \pm 0.004 \rm \; M_{sun}$ and $R_{sec} = 0.211 \pm 0.001 \rm \; R_{sun}$ as preliminary estimates (the mass of the primary star is yet unknown and we assumed 1 $\rm M_{Sun}$). This object is under study and will help in constraining the properties and evolution models of M-dwarfs.

\section{CONCLUSION}

The ASTEP instruments performed continuous observations during six winters in total in the extreme conditions of Antarctica. Thousands of lightcurves were recorded and led to several scientific results. We quantified the limitations of the instruments using the quality of the data. The main limitations are the PSF size and variations, which are due for the most part to the ground-level turbulence layer as well as the turbulent plumes generated by the heating system. We also encountered issues with the camera shutter. The pointing and guiding precisions are satisfactory. We also highlighted improvements that are technically feasible and would significantly improve the performances of the instruments. Overall, the outcomes of the ASTEP project will be highly beneficial to future optical and near-infrared telescopes installed at Dome C or at other Antarctic sites.

% Acknowledgments
\acknowledgments
The Dunlap Institute is funded through an endowment established by the David Dunlap family and the University of Toronto.
The ASTEP project was funded by the Agence Nationale de la Recherche (ANR), the Institut National des Sciences de l'Univers (INSU), the Programme National de Plan\'etologie (PNP), and the Plan Pluri-Formation OPERA between the Observatoire de la C\^ote d'Azur and the Universit\'e de Nice-Sophia Antipolis. The entire logistics at Concordia is handled by the French Institut Paul-Emile Victor (IPEV) and the Italian Programma Nazionale di Ricerche in Antartide (PNRA).
This research made use of tools provided by Astrometry.net \cite{Lang2010}.

% References


\begin{thebibliography}{10}

\bibitem{Fressin2005}
{Fressin}, F., {Guillot}, T., {Bouchy}, F., {Erikson}, A., {Gay}, J.,
  {L{\'e}ger}, A., {Pont}, F., {Rauer}, H., {Rivet}, J.-P., and {Valbousquet},
  F., ``{Antarctica Search for Transiting Extrasolar Planets},'' in [{\em EAS
  Publications Series}{\nolinebreak\hspace{0.1em}]},  {Giard}, M., {Casoli},
  F., and {Paletou}, F., eds., {\em EAS Publications Series} {\bf 14},
  309--312 (2005).

\bibitem{Crouzet2010}
{Crouzet}, N., {Guillot}, T., {Agabi}, A., {Rivet}, J.-P., {Bondoux}, E.,
  {Challita}, Z., {Fante{\"i}-Caujolle}, Y., {Fressin}, F., {M{\'e}karnia}, D.,
  {Schmider}, F.-X., {Valbousquet}, F., {Blazit}, A., {Bonhomme}, S., {Abe},
  L., {Daban}, J.-B., {Gouvret}, C., {Fruth}, T., {Rauer}, H., {Erikson}, A.,
  {Barbieri}, M., {Aigrain}, S., and {Pont}, F., ``{ASTEP South: an Antarctic
  Search for Transiting ExoPlanets around the celestial south pole},'' {\em
  \aap}~{\bf 511},  A36 (Feb. 2010).

\bibitem{Daban2010}
{Daban}, J.-B., {Gouvret}, C., {Guillot}, T., {Agabi}, A., {Crouzet}, N.,
  {Rivet}, J.-P., {Mekarnia}, D., {Abe}, L., {Bondoux}, E.,
  {Fante{\"i}-Caujolle}, Y., {Fressin}, F., {Schmider}, F.-X., {Valbousquet},
  F., {Blanc}, P.-E., {Le van Suu}, A., {Rauer}, H., {Erikson}, A., {Pont}, F.,
  and {Aigrain}, S., ``{ASTEP 400: a telescope designed for exoplanet transit
  detection from Dome C, Antarctica},'' in [{\em Ground-based and Airborne
  Telescopes III}{\nolinebreak\hspace{0.1em}]},  {\em \procspie} {\bf 7733},
  77334T (July 2010).

\bibitem{Fruth2014}
{Fruth}, T., {Cabrera}, J., {Csizmadia}, S., {Dreyer}, C., {Eigm{\"u}ller}, P.,
  {Erikson}, A., {Kabath}, P., {Pasternacki}, T., {Rauer}, H., {Titz-Weider},
  R., {Abe}, L., {Agabi}, A., {Gon{\c c}alves}, I., {Guillot}, T.,
  {M{\'e}karnia}, D., {Rivet}, J.-P., {Crouzet}, N., {Chini}, R., {Lemke}, R.,
  and {Murphy}, M., ``{Transit Search from Antarctica and Chile: Comparison and
  Combination},'' {\em \pasp}~{\bf 126},  227--242 (Mar. 2014).

\bibitem{Guillot2015}
{Guillot}, T., {Abe}, L., {Agabi}, A., {Rivet}, J.-P., {Daban}, J.-B.,
  {M{\'e}karnia}, D., {Aristidi}, E., {Schmider}, F.-X., {Crouzet}, N., {Gon{\c
  c}alves}, I., {Gouvret}, C., {Ottogalli}, S., {Faradji}, H., {Blanc}, P.-E.,
  {Bondoux}, E., and {Valbousquet}, F., ``{Thermalizing a telescope in
  Antarctica - analysis of ASTEP observations},'' {\em Astronomische
  Nachrichten}~{\bf 336},  638 (Sept. 2015).

\bibitem{Aristidi2009}
{Aristidi}, E., {Fossat}, E., {Agabi}, A., {M{\'e}karnia}, D., {Jeanneaux}, F.,
  {Bondoux}, E., {Challita}, Z., {Ziad}, A., {Vernin}, J., and {Trinquet}, H.,
  ``{Dome C site testing: surface layer, free atmosphere seeing, and
  isoplanatic angle statistics},'' {\em \aap}~{\bf 499},  955--965 (June 2009).

\bibitem{Fossat2010}
{Fossat}, E., {Aristidi}, E., {Agabi}, A., {Bondoux}, E., {Challita}, Z.,
  {Jeanneaux}, F., and {M{\'e}karnia}, D., ``{Typical duration of good seeing
  sequences at Concordia},'' {\em \aap}~{\bf 517},  A69 (July 2010).

\bibitem{Aristidi2013}
{Aristidi}, E., {Agabi}, A., {Fossat}, E., {Ziad}, A., {Abe}, L., {Bondoux},
  E., {Bouchez}, G., {Challita}, Z., {Jeanneaux}, F., {M{\'e}karnia}, D.,
  {Petermann}, D., and {Pouzenc}, C., ``{Dome C site testing: long term
  statistics of integrated optical turbulence parameters at ground level},'' in
  [{\em Astrophysics from Antarctica}{\nolinebreak\hspace{0.1em}]},  {Burton},
  M.~G., {Cui}, X., and {Tothill}, N.~F.~H., eds., {\em IAU Symposium} {\bf
  288},  300--301 (Jan. 2013).

\bibitem{Petenko2014}
{Petenko}, I., {Argentini}, S., {Pietroni}, I., {Viola}, A., {Mastrantonio},
  G., {Casasanta}, G., {Aristidi}, E., {Bouchez}, G., {Agabi}, A., and
  {Bondoux}, E., ``{Observations of optically active turbulence in the
  planetary boundary layer by sodar at the Concordia astronomical observatory,
  Dome C, Antarctica},'' {\em \aap}~{\bf 568},  A44 (Aug. 2014).

\bibitem{Lang2010}
{Lang}, D., {Hogg}, D.~W., {Mierle}, K., {Blanton}, M., and {Roweis}, S.,
  ``{Astrometry.net: Blind Astrometric Calibration of Arbitrary Astronomical
  Images},'' {\em \aj}~{\bf 139},  1782--1800 (May 2010).

\bibitem{Mekarnia2016}
{M\'ekarnia}, D., {Guillot}, T., {Rivet}, J.-P., {Schmider}, F.-X., {Abe}, L.,
  {Gonc\c{c}alves}, I., {Agabi}, A., {Crouzet}, N., {Fruth}, T., {Barbieri},
  M., {Bayliss}, D., {Zhou}, G., {Aristidi}, E., {Szulagyi}, J., {Daban},
  J.-B., {Fantei\"{i}-Caujolle}, Y., {Gouvret}, C., {Erikson}, A., {Rauer}, H.,
  {Bouchy}, F., {Gerakis}, J., and {Bouchez}, G., ``{Transiting planet
  candidates with ASTEP 400 at Dome C, Antarctica},'' {\em \mnras}  (in rev.).

\bibitem{Abe2013}
{Abe}, L., {Gon{\c c}alves}, I., {Agabi}, A., {Alapini}, A., {Guillot}, T.,
  {M{\'e}karnia}, D., {Rivet}, J.-P., {Schmider}, F.-X., {Crouzet}, N.,
  {Fortney}, J., {Pont}, F., {Barbieri}, M., {Daban}, J.-B.,
  {Fante{\"i}-Caujolle}, Y., {Gouvret}, C., {Bresson}, Y., {Roussel}, A.,
  {Bonhomme}, S., {Robini}, A., {Dugu{\'e}}, M., {Bondoux}, E., {P{\'e}ron},
  S., {Petit}, P.-Y., {Szul{\'a}gyi}, J., {Fruth}, T., {Erikson}, A., {Rauer},
  H., {Fressin}, F., {Valbousquet}, F., {Blanc}, P.-E., {Le van Suu}, A., and
  {Aigrain}, S., ``{The secondary eclipses of WASP-19b as seen by the ASTEP 400
  telescope from Antarctica},'' {\em \aap}~{\bf 553},  A49 (May 2013).

\bibitem{Chapellier2016}
{Chapellier}, E., {M\'ekarnia}, D., {Abe}, L., {Guillot}, T., {Agabi}, A.,
  {Rivet}, J.-P., {Schmider}, F.-X., {Crouzet}, N., and {Aristidi}, E., ``{A
  catalogue of eclipsing binaries and variable stars observed with ASTEP 400
  from Dome C, Antarctica},'' (in prep.).

\end{thebibliography}
\end{document}